\documentclass[twocolumn,aps,amssymb,prl]{revtex4}
\usepackage{amssymb}
\usepackage{graphicx}
\usepackage{epsfig,amsmath}

\begin{document}

\title{Single-walled carbon nanotube weak links: from Fabry-P\'{e}rot
to Kondo regime}

\author{F.~Wu$^{1}$, R.~Danneau$^{1}$, P. Queipo$^{2}$,
E.~Kauppinen$^{2}$, T.~Tsuneta$^{1}$, and P.~J.~Hakonen$^1$}
\affiliation{
$^1$Low~Temperature~Laboratory,~Helsinki~University~of~Technology,
Espoo, Finland \\
$^2$Center for New
Materials,~Helsinki~University~of~Technology,~Espoo,~Finland}

\date{\today} 

\begin{abstract}
We have investigated proximity-induced supercurrents in
single-walled carbon nanotubes in the Kondo regime and compared them
with supercurrents obtained on the same tube with Fabry-P\'{e}rot
resonances. Our data display a wide distribution of Kondo
temperatures \emph{$T_K$} = 1 - 14 K, and the measured critical
current $I_{CM}$ vs. \emph{$T_K$} displays two distinct branches;
these branches, distinguished by zero-bias splitting of the
normal-state Kondo conductance peak, differ by an order of magnitude
at large values of $T_K$. Evidence for renormalization of Andreev
levels in Kondo regime is also found.

\end{abstract}
\pacs{PACS numbers: 73.63.Fg, 74.25.Fy} \bigskip

\maketitle


An odd, unpaired electron in a strongly coupled quantum dot makes
the dot to behave as a magnetic impurity screened by delocalized
electrons. Such a Kondo impurity creates a peak in the density of
states at the Fermi level, thereby leading to characteristic Kondo
resonances with enhanced conductance around zero bias, which has
been observed in various quantum dot systems during the recent years
\cite{Goldhaber1998,Cronenwett1998,Nygard2000}. By studying low-bias
transport and multiple Andreev reflections (MAR) in multi-walled
carbon nanotubes (MWNT) contacted by superconducting leads, it has
been demonstrated that the Kondo resonances survive the
superconductivity of the leads when the Kondo temperature $T_{K}$
exceeds the superconducting gap $\Delta$ \cite{Buitelaar2002}; thus,
intricate interplay between Kondo behavior and superconductivity can
be studied in nanotube quantum dots.

In quantum dots made out of single walled carbon nanotubes (SWNT),
contacts play a crucial role in their transport properties: in the
highly transparent regime, Fabry-P\'{e}rot (FP) interference
patterns in the differential conductance can be observed
\cite{Liang2001}, whereas in the less transparent case, Coulomb
blockade peaks occur \cite{Tans1997}. In the intermediate regime,
zero bias conductance peaks alternate with Coulomb blockaded
valleys, highlighting Kondo resonances below Kondo temperature
$T_{K}$ due to odd numbers of spin in the cotunnelling process
between the dot and the leads \cite{Nygard2000}.

Gate-controlled, proximity-induced supercurrent has been reported
both in SWNTs \cite{Kasumov2paper, Morpurgo1999, JHerrero2006,
Cleuziou2006, Jorgensen2007, Zhang2007} and in MWNTs
\cite{Tsuneta2007, Pallechi2008}. Reasonable agreement with resonant
quantum dot weak link theories \cite{Beenakker1992theory} has been
reached in best of the samples (see e.g. Ref.
\onlinecite{JHerrero2006}).
In some of the experiments, Kondo-restored supercurrents were found
\cite{GRasmussen2007, Jorgensen2007} in otherwise Coulomb blockaded
Josephson junction case. In addition, when $T_K < \Delta$, $\pi$
Josephson junctions have been observed \cite{Cleuziou2006,
Jorgensen2007}.



Here we report a study of gate-tunable proximity-induced
supercurrents of an individual SWNT. We compare supercurrents in
Fabry-P\'{e}rot and Kondo regimes at the same normal state
conductance, and find smaller critical currents in the Kondo regime
up to $T_K \sim 10 \Delta$. In addition, we find that not just $T_K$
but also the shape of the Kondo resonance conductance peak affects
the magnitude of the supercurrent: resonances with zero-bias
splitting, which appear in about every second of our Kondo peaks,
result in a smaller critical current than for the regular Kondo
maxima.



Our nanotube samples were made using surface CVD growth with Fe
catalyst directly on oxidized, heavily-doped SiO$_2$/Si wafer. The
electrically conducting substrate works as a back gate, separated
from the sample by 150 nm of SiO$_2$. A sample with $L=0.7$ $\mu$m
length and $\phi=2$ nm diameter was located using an atomic force
microscope and the contacts on the SWNT were made using standard
e-beam overlay lithography. For the contacts, 10 nm of Ti was first
evaporated, followed by 70 nm of Al, in order to facilitate
proximity-induced superconductivity in Ti. Last, 5 nm of Ti was
deposited to prevent the Al layer from oxidation. The width of the
two contacts was $200$ nm and the separation between the them was
0.3 $\mu$m.

The measurement leads were filtered using an RC filter with time
constant of 10 $\mu$s at 1 K, followed by twisted pairs with tight,
grounded electrical shields for filtering between the still and the
mixing chamber, while the final section was provided by a 0.7-m long
Thermocoax cable on the sample holder.
In the measurements, differential conductance $G_d = dI/dV$ was
recorded using standard lock-in techniques. 
Voltage bias was imposed via a room-temperature voltage divider. The
normal state data were obtained by applying a magnetic field of $B
\sim 70$ mT perpendicular to the nanotube. The superconducting gap
of the contact material was found to be $\Delta_{g} = 125~\mu$eV,
and gate capacitance $C_g = 1.6$ aF was estimated from the measured
gate period of 0.1 V.

\begin{figure}
  \includegraphics[width=7cm]{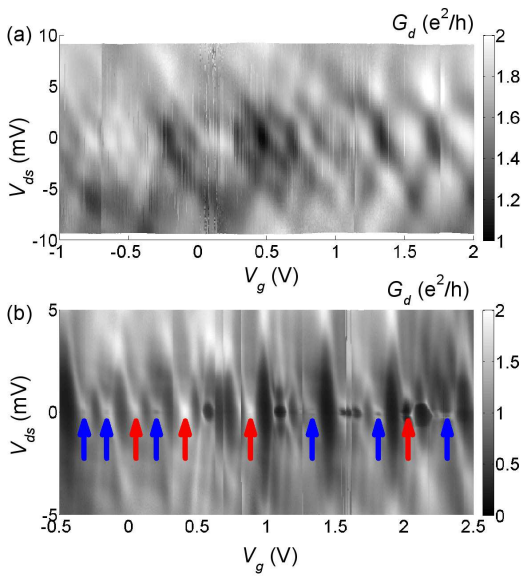}
     \caption{(Color online) Normal state differential conductance $G_d$ on
     the plane spanned by bias voltage $V_{ds}$ and gate voltage
     $V_g$ in (a) Fabry-P\'{e}rot regime, and (b) Kondo regime both at
     $T = 30$ mK.
     Normal states were achieved in all the cases with a magnetic
     field of $B = 70$ mT.
     Red and blue arrows in (b) refer to two types of
     resonance peaks, which have one magnitude difference in $I_{CM}$
     with similar Kondo temperature $T_K$. See text for more details.
     } \label{Gd-3D}
\end{figure}

The data presented in this paper have been measured in several cool
downs, thermal cycles, that have changed the contact conditions on
our sample. In the first cool down, the sample showed a strongly
asymmetric Fabry-P\'{e}rot pattern with one low-transmission
(spin-degenerate) channel and another one with high transmission;
the zero-bias conductance was limited to $2e^2/h$ as a consequence
\cite{Wu2007}. A scan of differential conductance $G_d(V_{ds}, V_g)$
versus bias voltage $V_{ds}$ and gate voltage $V_g$ is shown in Fig.
\ref{Gd-3D}(a) at $B \sim 70$ mT. In the absence of magnetic field,
a gate-voltage-dependent supercurrent is observed in the SWNT. The
measured critical supercurrent $I_{CM}$ varies periodically with the
gate voltage $V_g$, reaching a maximum of 4.8 nA at zero bias normal
state conductance $G_N = G_d|_{V_{ds}=0} = 2.03 e^2/h$. The
$I_{CM}R_N$ product is $V_g$-dependent and it changes in a similar
fashion as $I_{CM}$ and the inverse of the normal state resistance
$G_N$. This result is
similar to what has been observed in a superconducting SWNT in
Fabry-P\'{e}rot regime \cite{JHerrero2006}.


After a few thermal cycles, the transport of SWNT changed from
Fabry-P\'{e}rot into Kondo type of behavior as seen in Fig.
\ref{Gd-3D}(b). The $G_d$ map displays a series of Coulomb blockade
diamonds (even number of electrons) alternating with Kondo ridges,
marked by the arrows (odd number of electrons). The Kondo ridges are
rather wide and the Kondo temperatures, which are deduced from the
half width at half maximum of the resonant conductance peaks versus
bias voltage $G_d(V_{ds})$ (see below), range over $T_K=1 - 14$ K.
We find that both the critical current and zero-bias conductance are
smaller compared with Fabry-P\'{e}rot regime, even in the Kondo
resonances with the highest $T_K$.

\begin{figure*}
  \includegraphics[width=16cm]{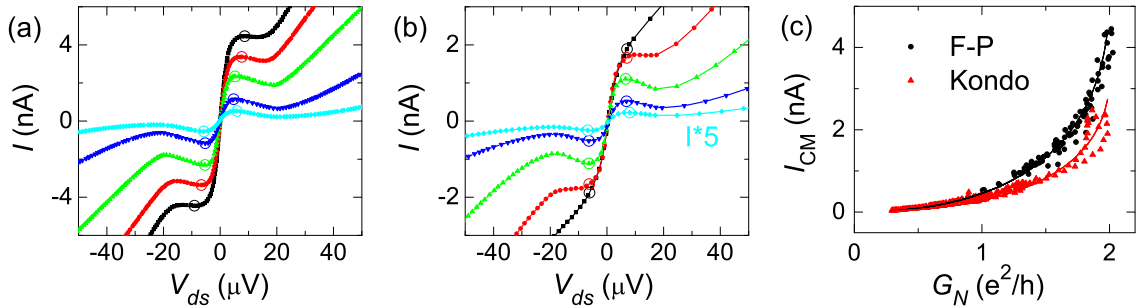}
     \caption{(Color online) Superconducting $I-V$ curves at a few
     gate voltage values in (a) Fabry-P\'{e}rot regime, and (b) Kondo
     regime. The circles with different colors show how the measured critical
     current $I_{CM}$ was determined \cite{noteICM}.
     $I_{CM}$ versus zero-bias normal state conductance $G_N$,
     measured for a resonance with $T_K = 14$ K,
     is displayed in (c) where the black dots and red triangles refer to
     Fabry-P\'{e}rot (several resonances) and Kondo data, respectively.
     Data in (a) were measured in the same cool down as Fig. \ref{Gd-3D}(a)
     at $T = 60$ mK; data in (b) were taken from
     another cool down after Fig. \ref{Gd-3D}(b) at $T = 60$ mK
     with unchanged $G_N$. The current of the smallest $I_{CM}$
     curve in (b) has been amplified by a factor of 5 for
     clarity. Black and red solid lines in (c) are theoretical
     fits using Eq. (\ref{Ic-Gn_formu}).
     } \label{IV_SC}
\end{figure*}

The superconducting state IV curves in both Fabry-P\'{e}rot and
Kondo regimes are shown in Fig. \ref{IV_SC}. As the sample is
voltage biased, negative differential resistance (NDR) is observable
in Fabry-P\'{e}rot regime. However, in Kondo regime, NDR  occurs
only at small measured critical current $I_{CM}$ and it disappears
around the maximum of the Kondo resonance peak where $I_{CM}$ is
large.
We note that zero bias resistance and the IV curves evolve
smoothly with $V_g$ around the Kondo resonance without any sudden
jumps, and that $T_K > \Delta_g$. We ascribe the disappearance of
NDR to the presence of large MAR-induced subgap current, which is
stronger with respect to the supercurrent in the Kondo regime than
in the FP case.

The nanotube together with superconducting leads can be considered
as a resonant level quantum dot, and thus the two-barrier
Breit-Wigner model is applicable to model the behavior
\cite{Beenakker1992theory}. In our case, the measured $I_{CM}$ is
nearly one order of magnitude smaller than the theoretical
prediction $I_0 = e\Delta_g/\hbar \thickapprox 30$ nA with one
resonant spin-degenerate level \cite{notelength}. Taking into
account the phase diffusion in an underdamped, voltage-biased
Josephson junction \cite{Ingold1994}, the measured $I_{CM} \propto
{E_J}^2 \propto {I_C}^2$. With Breit-Wigner model for wide resonance
limit $h\Gamma
>> \Delta_g$ and transmission probability $\alpha_{BW}$, we have $I_C =
I_0[1-(1-\alpha_{BW})^{1/2}]$, so the $I_{CM} - G_N$ relation can be
written as
\begin{equation}\label{Ic-Gn_formu}
     I_{CM} = I_{0M} [ 1- (1 - \frac{1}{2}g_n)^{\frac{1}{2}}]^2,
\end{equation}
where $I_{0M}$ denotes the maximum measurable critical current when
the scaled conductance $g_n = G_N/(e^2/h) \rightarrow 2$.
This equation is written for one spin-degenerate channel (the Kondo
case) where the transmission coefficient is obtained from
$\frac{1}{2}g_n$, and the prefactor depends on $T_K$; in our case it
also applies approximately to the asymmetric FP conduction as one of
the spin-degenerate transmission channels is greatly suppressed
\cite{Wu2007}.
The fit of Eq. (\ref{Ic-Gn_formu}) to our data is displayed in Fig.
\ref{IV_SC} (c), with $I_{0M} = 5.3$ nA and $3.3$ nA corresponding
to Fabry-P\'{e}rot and Kondo regimes respectively (the latter at
$T_K = 14$ K).

\begin{figure}
  \includegraphics[width=7cm]{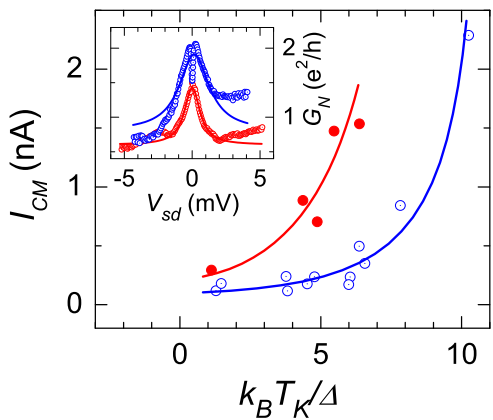}
     \caption{(Color online) Measured critical current $I_{CM}$ versus
     scaled Kondo temperature $k_BT_K/\Delta$ for Kondo resonances
     marked by red and blue arrows
     in Fig. \ref{Gd-3D}.
     Peaks with zero-bias splitting are denoted by
     blue circles,
     while red dots refer
     to non-split peaks in Fig. \ref{Gd-3D}(b).
     The solid red and blue curves are to guide the eyes.
     The inset shows two typical $G_N - V_{ds}$ relations
     for the different kinds of conductance peaks and their Lorentzian
     fits. The curve for non-split peak has been shifted downwards by
     0.3 units for clarity.} \label{Ic-TK}
\end{figure}

As in the FP regime, the largest critical current over a Kondo
resonance corresponds to the peak value of the normal state
conductance. In addition, the magnitude of $I_{CM}$ depends on the
width of the resonance in bias voltage, i.e. on $T_K$. We have
fitted the conductance  peaks $G_d(V_{ds})$ with a Lorentzian
function in order to extract the Kondo temperature $T_K$. The
resulting $I_{CM} - T_K$ correlation is plotted in Fig. \ref{Ic-TK}
which displays two branches, instead of a single-valued correlation
as observed by Grove-Rasmussen \emph{et al.} \cite{GRasmussen2007}.
The upper and lower branches involve the resonance peaks marked in
Fig. \ref{Gd-3D} by red and blue arrows, respectively. Due to the
problem of trapped charge fluctuating on the back gate, we have been
forced to present only data on which we are sure of the
identification between critical current and normal state
conductance.
As shown in the inset of Fig. \ref{Ic-TK}, in the data of the lower
branch, there is a small dip on the zero-bias conductance peak
signifying zero-field splitting of the Kondo resonances marked by
blue arrows. The Lorentzian fitting on the split peaks is somewhat
approximative, and the fitted $T_K$ remains a bit smaller than from
the true half width. Nevertheless, this uncertainty is insignificant
on the scale of separation of the upper and lower branches in Fig.
\ref{Ic-TK}.
Notice that $k_BT_K \sim E_C > \Delta >> k_BT$ is valid for all of
the measured resonance peaks, which indicates that the double branch
structure originates from the competition between Kondo effect and
the Coulomb blockade.

Zero-field splitting seems to take place in our data in every second
Kondo resonance, as seen in the nearly alternating sequence of red
and blue arrows in Fig. \ref{Gd-3D} (b). Zero-field Kondo-peak
splitting has previously been reported in Ref.
\onlinecite{marcus2004}, where the the splitting originates from
magnetic impurity, which is different from our case as the splitting
should then be seen at every Kondo resonance. Using the standard
fourfold shell-filling sequence, it is hard to explain our findings.
Split Kondo ridges may be observable when the dot is occupied by two
electrons ($N$ = 2) \cite{Nygard2000, Babic2004}, and the energy
scale of the splitting equals to the gap between singlet ground
state and triplet excited state. This, however, should be bordered
from both sides by standard spin-half Kondo peaks, a sequence that
we cannot identify in our data.
From the normal state bias maps, the characteristic zero-bias
splitting energy can be estimated as $\Delta_{ZBS} \thicksim 0.4$
meV, which is well above MAR peak of superconducting electrodes
$2\Delta_g = 0.25$~meV and the typical singlet-triplet excitation
energy as found in Ref. \onlinecite{Nygard2000}. We conjecture that
the observed zero-field splitting is related to the SU(4) Kondo
effect which is peculiar to carbon nanotubes \cite{JHerrero2005,
Makarovski2007} and which has been shown to lead to a dip in the
density of states at small energies \cite{Lim2006}. Alternatively,
zero-field splitting may be related with the recent observation of
non-negligible spin-orbit coupling in SWNTs \cite{Kuemmeth2008}. In
any case, SU(4) Kondo can explain the unusually high $T_K$ by the
enhanced degeneracy of a multiple-level quantum dot
\cite{Inoshita1993}.

\begin{figure}
  \includegraphics[width=8cm]{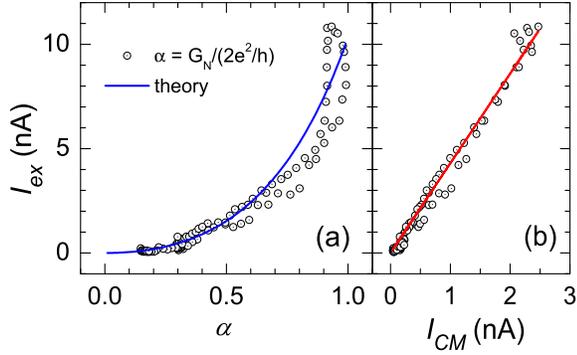}
     \caption{(Color online) Excess current $I_{ex}$
     of one Kondo resonance with $T_K = 14$ K
     at $T = 90$ mK versus
     (a) transmission coefficient
     $\alpha = G_N/(2e^2/h)$ and (b) measured critical current $I_{CM}$.
     The blue line in (a) is the theoretical curve from Eq.
     (\ref{Iex_formu}) with $\tilde{\Delta} = 100\mu$eV,
     and red line in (b) gives linear fit of $I_{ex} / I_{CM} = 4.3$ .} \label{Iex-a}
\end{figure}

According to theory \cite{Yeyati2003}, the width of Andreev levels
can be substantially renormalized by the Kondo effect, which would
also modify the IV curve.
In order to look for the gap renormalization, we have extracted
the excess current $I_{ex}$ as a function of normal state
transmission coefficient $\alpha$, which is displayed in Fig.
\ref{Iex-a}, with $\alpha$ calculated from $\alpha = G_N /
(2e^2/h)$, and $I_{ex}$ determined by the difference of integration
from $G_d - V_{ds}$ curves in superconducting/normal state like
in Ref. \onlinecite{Jorgensen2006}.
The relation between $I_{ex}$ and $\alpha$ in a quantum point
contact \cite{Averin1995, Bratus1995, Cuevas1996} can be written as
$I_{ex} = I_{ex1} + I_{ex2}$, where
\begin{eqnarray}\label{Iex_formu}
     I_{ex1} = \frac{e\Delta}{h} \frac{\alpha^2}{(2-\alpha)\sqrt{1-\alpha}}
     \ln \left[ \frac{1+ [2\sqrt{1-\alpha}/(2-\alpha)] } {1-
     [2\sqrt{1-\alpha}/(2-\alpha)]} \right], \nonumber\\
     I_{ex2} = \frac{e\Delta}{h} \alpha^2 \left[
     \frac{1}{1-\alpha}+\frac{2-\alpha}{2(1-\alpha)^{3/2}}
     \ln \left( \frac{1-\sqrt{1-\alpha}}{1+\sqrt{1-\alpha}} \right)
     \right].
\end{eqnarray}
The blue curve in Fig.~\ref{Iex-a} (a) illustrates
Eq.~\ref{Iex_formu} with $\Delta \equiv \tilde{\Delta} = 100 \mu$eV,
indicating a gap renormalization of $\tilde{\Delta}/\Delta_g = 0.8$.
We have also investigated the relation between $I_{ex}$ and $I_{CM}$
at different gate voltages. The data is shown in Fig. \ref{Iex-a}
(b), which yields a linear relation with $I_{ex}/I_{CM}=4.3$ at $T_K
= 14$ K; this arises because both are proportional to $\alpha^2$.
MAR-induced current at large bias voltage gives $I_{AR} =
4e\Delta/h$ \cite{Avishai2003}. By taking into account that $I_{0M}
\sim \frac{1}{10}I_0$, we get $I_{AR}/I_{CM} \sim \frac{20}{\pi}$,
which is close to the measured $I_{ex}/I_{CM}$ value.


In summary, we have investigated experimentally the
proximity-effect-induced supercurrents in SWNTs in the Kondo regime
and compared them with results in the Fabry-Perot regime with
equivalent conductance. In the Kondo regime, two different types of
resonances, either split or non-split at zero-bias, were observed
and this behavior reflected also in the magnitude of supercurrent
that displayed two branches vs. $T_K$. The excess current in Kondo
regime was analyzed using MAR theory and renormalization of Andreev
levels by 80 \% was obtained.

We wish to acknowledge fruitful discussions with S. Andresen,
J.~C.~Cuevas, T. Heikkil\"a, J. Voutilainen, A. D. Zaikin, K.
Flensberg, G. Cuniberti, T.~Kontos, L.~Lechner, P.-E. Lindelof,
C.~Strunk and P. Virtanen. This work was supported by the Academy of
Finland and by the EU contract FP6-IST-021285-2.


\begin{thebibliography}{99}


\bibitem{Goldhaber1998} D. Goldhaber-Gordon \emph{et al.}, Nature {\bf
391}, 156 (1998).

\bibitem{Cronenwett1998} S. M. Cronenwett , T. H. Oosterkamp, and L. P.
Kouwenhoven, Science {\bf 281}, 540 (1998).

\bibitem{Nygard2000} J. Nyg{\aa}rd, D. H. Cobden, and P. E. Lindelof,
Nature \textbf{408}, 342 (2000).

\bibitem{Buitelaar2002} M. R. Buitelaar, T. Nussbaumer,
and C. Sch\"{o}nenberger, Phys. Rev. Lett. \textbf{89}, 256801
(2002).

\bibitem{Liang2001} W. Liang \emph{et al.}, Nature \textbf{411}, 665 (2001).

\bibitem{Tans1997} S. J. Tans  \emph{et al.}, Nature \textbf{386}, 474 (1997).










\bibitem{Kasumov2paper} A. Yu. Kasumov  \emph{et al.},
Science \textbf{284}, 1508 (1999); A. Yu. Kasumov \emph{et al.},
Phys. Rev. B \textbf{68}, 214521 (2003).

\bibitem{Morpurgo1999} A. F. Morpurgo  \emph{et al.},
Science \textbf{286}, 263 (1999).

\bibitem{JHerrero2006} P. Jarillo-Herrero, J. A. van Dam, and L. P.
Kouwenhoven, Nature \textbf{439},953 (2006).

\bibitem{Cleuziou2006} J.-P. Cleuziou  \emph{et al.}, Nature Nanotech.
\textbf{1}, 53 (2006).

\bibitem{Jorgensen2007} H. I. J{\o}rgensen  \emph{et al.}, Nano Lett.
\textbf{7}, 2441 (2007).


\bibitem{Zhang2007} Y. Zhang, G. Liu, and C. N. Lau, Nano Res. \textbf{1}, 145
(2008).

\bibitem{Tsuneta2007} T. Tsuneta, L. Lechner, and P. J. Hakonen,
Phys. Rev. Lett. \textbf{98}, 087002 (2007).

\bibitem{Pallechi2008} E. Pallecchi  \emph{et al.},
Appl. Phys. Lett. \textbf{93}, 072501 (2008).

\bibitem{Beenakker1992theory} See, \emph{e.g.}, C. W. J. Beenakker and
H. van Houten, Single-electron Tunneling and Mesoscopic Devices (eds
H. Koch and H. L\"{u}bbig) (reprinted in arXiv:cond-mat/0111505)
175--179 (Springer, Berlin, 1992).

\bibitem{GRasmussen2007} K. Grove-Rasmussen, H. I. J{\o}rgensen,
and P. E. Lindelof, New J. Phys. \textbf{9}, 124 (2007).

\bibitem{Wu2007} F. Wu  \emph{et al.}, Phys. Rev. Lett. \textbf{99},
156803 (2007).

\bibitem{notelength} This value does not include the
effect of finite length of the nanotube, which would suppress the
maximum $I_0$ a bit. See, \emph{e.g.}, A. V. Galaktionov and A. D.
Zaikin, Phys. Rev. B \textbf{65}, 184507 (2002).

\bibitem{Ingold1994} G.-L. Ingold, H. Grabert, and U. Eberhardt,
Phys. Rev. B \textbf{50}, 395 (1994).






\bibitem{noteICM} When NDR
exists, we extract $I_{CM}$ from the local current maximum
$I_{CM_p}$ and minimum $I_{CM_n}$ using $I_{CM} = (I_{CM_p} -
I_{CM_n})/2$. Around the Kondo resonance peaks without NDR, $I_{CM}$
is obtained using the averaged voltages, $\overline{V_{CM_p}}$ and
$\overline{V_{CM_n}}$, respectively, of the $I_{CM_p}$ and
$I_{CM_n}$ peak positions at lower $I_{CM}$, and taking $I_{CM} =
(I(\overline{V_{CM_p}}) - I(\overline{V_{CM_n}})/2$.

\bibitem{marcus2004} J. Nyg{\aa}rd  \emph{et al.}, arXiv:cond-mat/0410467v2.

\bibitem{Babic2004} B. Babi\'{c}, T. Kontos, and C.
Sch\"{o}nenberger, Phys. Rev. B \textbf{70}, 235419 (2004).

\bibitem{JHerrero2005} P. Jarillo-Herrero  \emph{et al.}, Nature
\textbf{434}, 484 (2005).

\bibitem{Makarovski2007} A. Makarovski  \emph{et al.},
Phys. Rev. B \textbf{75}, 241407(R) (2007).

\bibitem{Lim2006} J. S. Lim  \emph{et al.},
Phys. Rev. B \textbf{74}, 205119 (2006).

\bibitem{Kuemmeth2008} F. Kuemmeth  \emph{et al.},
Nature \textbf{452}, 448 (2008).

\bibitem{Inoshita1993} T. Inoshita  \emph{et al.}, Phys. Rev. B
\textbf{48}, 14725 (1993).

\bibitem{Yeyati2003} A. Levy Yeyati, A. Mart\'{i}n-Rodero, and E.
Vecino, Phys. Rev. Lett. \textbf{91}, 266802 (2003).

\bibitem{Jorgensen2006} H. I. J{\o}rgensen  \emph{et al.}, Phys. Rev.
Lett. \textbf{96}, 207003 (2006).

\bibitem{Averin1995} D. Averin and A. Bardas,
Phys. Rev. Lett. \textbf{75}, 1831 (1995).

\bibitem{Bratus1995} E. N. Bratus, V. S. Shumeiko, and G. Wendin,
Phys. Rev. Lett. \textbf{74}, 2110 (1995).

\bibitem{Cuevas1996} J. C. Cuevas, A. Mart\'{i}n-Rodero, and A. Levy Yeyati,
Phys. Rev. B \textbf{54}, 7366 (1996).



\bibitem{Avishai2003} Y. Avishai, A. Golub and A. D. Zaikin,
Phys. Rev. B \textbf{67}, 041301(R) (2003).




\end{thebibliography}
\end{document}